\documentclass[twocolumn,showpacs,showkeys]{revtex4}

\usepackage{lineno,hyperref}
\usepackage{graphicx}
\usepackage{amssymb}
\usepackage{amsmath}
\usepackage{bm}
\usepackage{xcolor,pict2e}

\begin{document}
\setcounter{page}{1}
\title[]{Tuning the Stiffness Asymmetry of Optical Tweezers via Polarization Control}
\author{Jinmyoung \surname{So}}
\author{Jai-Min \surname{Choi}}
\email{jaiminchoi@jbnu.ac.kr}
\thanks{Fax: +82-63-270-2806}
\affiliation{Division of Science Education and Institute of Fusion Sciences, Chonbuk National University, Jeonju 54896, Korea}
\date[]{}

\begin{abstract}
Optical tweezers that utilize a highly focused, linearly polarized laser beam exhibit strong trap stiffness asymmetry, which originates from the anisotropic field distribution in the transverse plane. Based on the previous study of polarization-dependent focused field distribution, we explore its manifestation in optical trap in terms of trap stiffness asymmetry. Our results demonstrate that polarization control provides a versatile tuning knob for tailoring optical potential landscape even in case of a strongly modified focused field in the presence of dielectric spheres larger than the wavelength of a trap beam. 
\end{abstract}



\maketitle

\section{Introduction}
A tightly focused laser beam can help overcome radiation pressure, resulting in the stable optical trapping of microscopic particles in an optical potential; this principle is physically realized in instruments known as ``optical tweezer'' (OTs) \cite{Ashkin1970, Ashkin1986}. OTs provide a contactless optical clamping method and additional means of governing the system dynamics (e.g., studying interaction dynamics in colloidal systems \cite{Crocker1999,Trankle2012,Romodina2015} and microgyroscope applications \cite{Donato2014,Arita2013}). Moreover, by harnessing the extreme force sensitivity such methods have various applications in biophysics and precision control \cite{Grier2003}. In these applications, tailoring the landscape of the optical potential is of significant importance, and considerable efforts have been devoted to engineering the optical potential in order to realize the modeled systems desirable for manifesting novel characteristics of research interest (e.g., scanning-line OTs \cite{Rogers2014}).

Recent studies on this subject in the context of highly focused OTs have shown that strong asymmetry in trap stiffnesses \cite{Rohrbach2005,Zakharian2006,Madadi2012}, which is equivalent to asymmetric trap potential, is primarily due to the anisotropic electromagnetic (EM) field distribution around a trap region. Rohrbach {\it et al.} illustrated the trap stiffness asymmetry in their OT setup based on a water-immersion objective lens \cite{Rohrbach2005}. Zakharian {\it et al.} used the finite-difference time-domain (FDTD) method to investigate polarization effects on the optical potential formed in air and in water \cite{Zakharian2006}. Madadi {\it et al.} explored the aberration effect that is caused by the refractive index mismatch between the interfaces of the media involved; more specifically, they investigated the trap potential asymmetry that arises in OTs composed of an oil-immersion objective lens \cite{Madadi2012}. Maragò {\it et al.} elucidated the trap potential anisotropy of OT by analyzing the autocorrelation and cross-correlation of the center-of-mass and angular motion of optically trapped nanotube bundle \cite{Marago2008}, and Jinxin {\it et al.} illustrated the two-dimensional trapping-potential profile using the optical bottle method \cite{Fu2013}.

While the abovementioned studies \cite{Rohrbach2005,Zakharian2006,Madadi2012} focused on the characterization of trap stiffness asymmetry in OTs with a linearly polarized trap beam, we address the same subject with the aim of achieving a control parameter over the stiffness asymmetry by using the polarization state of the trapping beam. To investigate the effect of the polarization state of the trap laser beam on the trap potential asymmetry, we constructed a custom-built OT setup equipped with polarization control. The input beam is prepared with linear, elliptic, and circular polarization, and the resulting trap potential asymmetry is characterized in terms of the trap stiffnesses (spring constants) in the transverse plane. The trap stiffnesses are measured for each polarization state from the motion of trapped particles by using the back-focal-plane interferometry (BFPI) technique \cite{Gittes1998}. Theorerical estimation for the experimental observations is also presented based on the generalized Lorent-Mie theory, which is revised to encompass the general polarization state of a trapping beam \cite{Noh2015}.

\section{Experimental setup and preliminary measurement}
We constructed a custom-built OT setup equipped with polarization control, as shown in Fig. 1(a). The output of the diode laser system is fed to a tapered amplifier, resulting in 1.6 W of output power with a tuned wavelength ($\lambda_{0}$) of 780 nm. The power-amplified beam is guided to the experimental region through a single-mode polarization-maintaining fiber, yielding the transmitted power of about 250 mW. The first half-wave plate (HWP1) and polarizing beam splitter (PBS) are used to adjust the input beam power, whereas the second half-wave plate (HWP2) and the quarter-wave plate (QWP) define the polarization state. As benchmark conditions, the maximum power of the incident beam and the collimated beam size before the water-immersion objective lens (UPLSAPO 60XW, Olympus) were 150 mW and 4 mm (corresponding to a filling factor of ~1), respectively. Part of that beam power was reserved for power stabilization. The wave plates were initially aligned to produce linearly polarized light along the horizontal direction in the laboratory frame:$\vec{E}^{in}=E_{0}\hat{x}$  . A 170-$\mu$m-thick cover glass (NO1.5H, ZEISS) and a nominal slide glass were sandwiched together as sidewalls in order to construct the sample chamber using a 200-$\mu$m-thick, double-sided 3M tape. The polystyrene spheres (Bangs Laboratories) of diameter  300 $\pm$ 15 nm, with the index of refraction $n_{p}$ = 1.579 at $\lambda_{0}$ = 780 nm, were diluted and dispersed in deuterium oxide (D$_{2}$O) to minimize heat-induced convection. Additional procedures for hydrophobic coating of the glass surfaces and stabilizing surfactant were made following the protocol provided in Ref. \cite{Lee2007}. The condenser lens (L1) collects part of the trapping beam together with the scattered light. The superposed field pattern in the back focal plane of L1 was imaged onto the quadrant photodiode (QPD) via the relay lens L2, which is known as the BFPI technique \cite{Gittes1998}. An iris at the back focal plane of L1 was used to optimize the sensitivity of BFPI by adjusting the effective numerical aperture (NA) of the condenser lens \cite{Friedrich2012}, and the auxiliary components (LED, L3, and DM) are part of the imaging setup. The output signals ($\delta V_{x}$, $\delta V_{y}$, $V_{SUM}$) of the QPD were logged using a high-speed digital scope (5444B, PicoScope) with a 5-$\mu$s sampling time; each measurement was performed for 2 s.

\begin{figure}[t]
\centering
\includegraphics[width=\columnwidth]{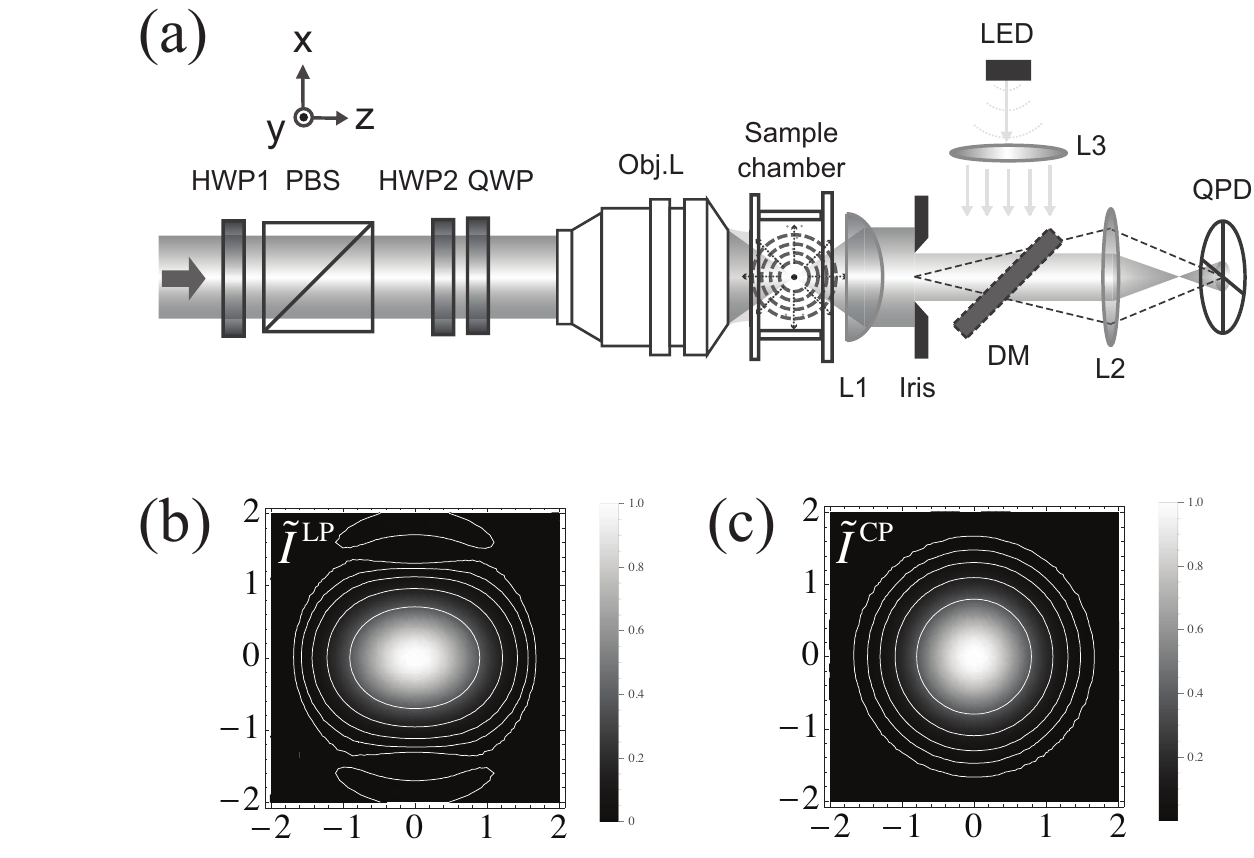}
\caption{(a) Schematic  of the experimental setup. HWP: half-wave plate. PBS: polarizing beam splitter. QWP: quarter-wave plate. Obj. L: a water-immersion objective lens. L1 and L2: condenser lens and relay lens, respectively. QPD: quadrant photodiode. LED: light-emitting diode. The dichroic mirror (DM) is removed during PSD measurements. (b) The normalized intensity distribution for a linearly polarized (LP) input beam at the focal plane. (c) The normalized intensity distribution for a circularly polarized (CP) input beam at the focal plane. In (b-c), the dashed white lines outline the ($e^{-1}$, $e^{-2}$, $e^{-3}$, $e^{-4}$, $e^{-5}$) levels and the frame ticks are given units of {\it w$_{0}$}.}\label{fig1}
\end{figure}

Having a diffraction-limited spot is crucial for our research because it ensures exclusion of other contributions, e.g., aberration effects, and trap asymmetry itself strongly depends on the spot size. We exploited the scattered light from the particle to produce a tight spot. For this diagnostic purpose, the input trap beam was prepared with a circular polarization, i.e.,$E_{0}\hat{x} \rightarrow E_{0}(\hat{x}+i\hat{y})/\sqrt{2}$, and part of the scattered light from the test sphere being trapped was recollected by the objective lens. It passed the QWP again, mostly resulting in the polarization that is rotated by 90$^{\circ}$: $E_{0}(\hat{x}-i\hat{y})/\sqrt{2} \rightarrow E_{0}\hat{y}$. The rotated scattered light was redirected by the PBS and mode-filtered through a 15-$\mu$m pinhole to eliminate the reflected light from the other interfaces, except for that of the trapped particle. We maximized the guided power by adjusting the trap beam alignment and the correction collar of the objective lens \cite{Reihani2011}.

Figures 1(b) and 1(c) show the numerically calculated intensity distribution based on vectorial diffraction theory considering the following experimental parameters (as was discussed in our previous study \cite{Noh2015} and references are therein): NA of the objective lens is 1.2, focal length f = 3 mm, filling factor = 1.0, vacuum wavelength $\lambda_{0}$ = 780 nm, and the refractive index ({\it n$_m$}) of D$_{2}$O is 1.324. The intensity distribution is markedly different depending on the input beam polarization states: an elongated distribution for linearly ($\hat{x}$) polarized beam (Fig. 1(b)) and an isotropic distribution for circularly polarized beam (Fig. 1(c)). The minimum spot size ({\it w$_{0}$}) was estimated to be 293 nm through a Gaussian fit of the numerically calculated field pattern \cite{Noh2015,Friedrich2012}, whereas the conventional definition of optical resolution (the radius of the primary Airy disk) gives the relevant value as 270 nm. Based on our experimental results, the beam waist ({\it w$'_{0}$}) is chosen to be 300 nm, which is 2$\%$ larger than the ideal case, although we elaborated on producing the tightest spot size as possible.

\begin{figure}[b]
\centering
\includegraphics[width=\columnwidth]{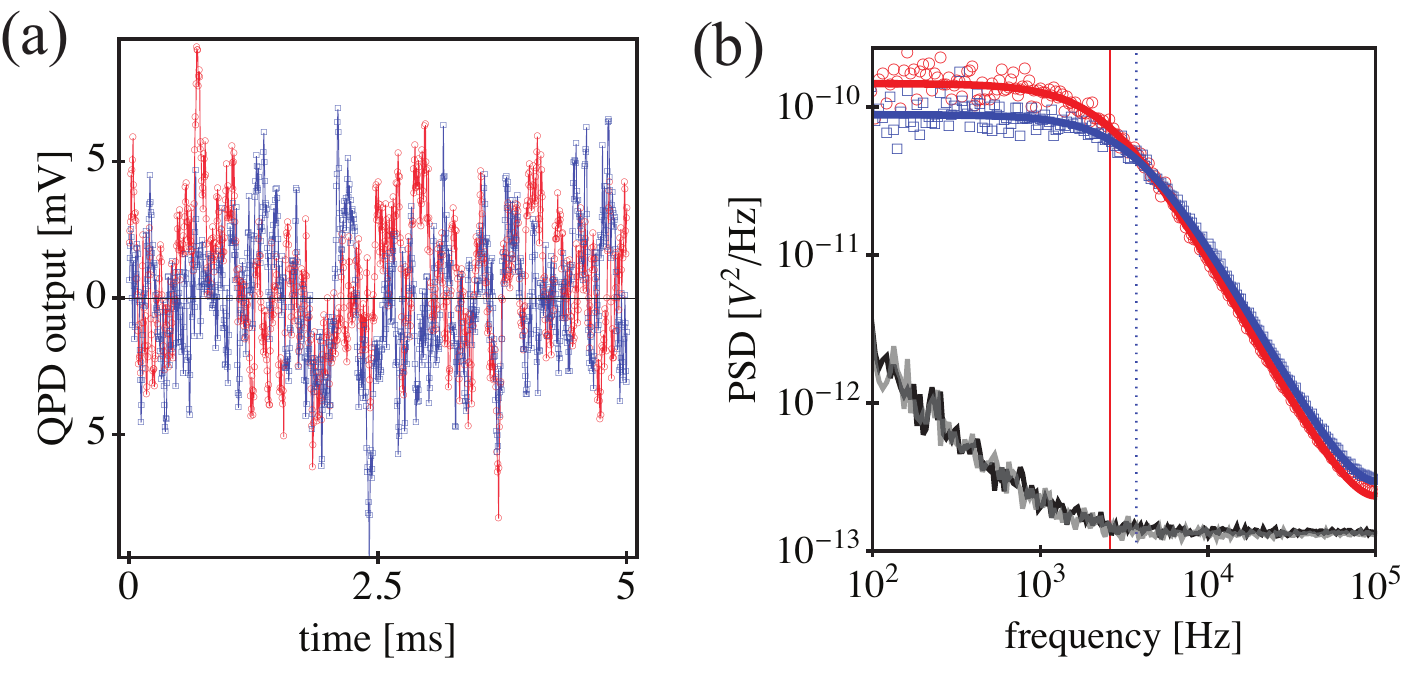}
\caption{(a) Part of QPD outputs: $\delta V_x$ (red) and $\delta V_y$ (blue). (b) PSD of QPD outputs: $\bigcirc$ (red) and $\square$ (blue) represent the PSD of $\delta V_x$ and $\delta V_y$, respectively. Black and gray lines represent the noise levels of OT. The two vertical lines indicate the corner frequencies: $f_{c,x}$ (red solid line) and $f_{c,y}$ (blue dotted line).}\label{fig2}
\end{figure}

Part of the $\delta V_x$ (red points) and $\delta V_y$ (blue points) data are presented in Fig. 2(a), and the corresponding power spectral density (PSD) is shown in Fig. 2(b) with the linearly ($\hat x$) polarized trap beam. The beam power in the trap region was estimated to be $P_{0}$ = 40 mW. The stochastic motion of the trapped particle can be modeled in the form of the Langevin equation with an additional Hookean force term, where the restoring force in a specific direction is characterized by the corresponding stiffness $k_i$, where $i \in (x, y, z)$. The power spectra for each degree of freedom can be approximated to the Lorentzian function $S_{i}(f)=(D/2\pi^{2})/(f^{2}_{c,i}+f^{2})$ \cite{Berg2004}, where {\it D} is the diffusion constant. The corner frequency $f_{c,i}$ is given by $k_{i}/2\pi \gamma$ and $\gamma$ is the frictional constant given by Stokes' law, $\gamma = 6\pi \eta a$. The dynamic viscosity ($\eta$) of D$_2$O is 1.12$\times$10$^{-3}$ kg/m/s and the nominal radius of the PS spheres is {\it a} = 150 nm. The thick solid lines (red and blue curves in Fig. 2(b)) represent the results of the Lorentzian fit following the procedures reported in Ref. \cite{Berg2004, Tolic2004}, which give the corner frequencies for each degrees of freedom in the transverse plane; ($f_{c,x}$, $f_{c,y}$) = (2497 $\pm$ 25, 3616 $\pm$ 33) Hz, respectively. The associated trap stiffnesses ($k_x$, $k_y$) are (49, 71) pN/$\mu$m and the corresponding stiffness asymmetry factor, $s_{T}=1-k_{x}/k_{y}$, in the transverse plane is $s^ \text {LP} _{T, \text {EXP}}$ = 0.31 for the linearly ($\hat{x}$) polarized trap beam. Our previous numerical study based on the revised generalized Lorenz-Mie theory (GLMT) predicted the transverse asymmetry factor $s^\text{LP}_{T, \text{GLMT}}$ = 0.3 under an aberrations-free condition \cite{Noh2015}, which is in a reasonable agreement with the experimentally measured value. 

\section{Characterization and control of stiffness asymmetry}
The landscape of the optical potential depends not only on the EM filed distribution in the trapping region, but also on the characteristics of the particle being trapped, i.e., its size, shape, and index of refraction, and the index of refraction of the suspension medium as was demonstrated in the previous studies  \cite{Rohrbach2005,Zakharian2006,Madadi2012,Noh2015}. We focused on the systematic investigation of the polarization effect on the trap potential asymmetry using the polystyrene (PS) spheres with the diameters of 300 $\pm$ 15 nm ($< \lambda_{0}$) and 1000 $\pm$ 30 nm ($> \lambda_{0}$). Polarization-induced stiffness asymmetry was investigated as a function of the prepared input polarization state. The experimental parameters used in the preliminary experiment are used here again. 

The angular variation of the trap stiffnesses $k_x$ and $k_y$ are estimated from PSD measurements for the linearly polarized input beam rotated by $\theta_R$ using HWP2, where $k_x$ and $k_y$ are stiffnesses along the horizontal ($\hat{x}$) and vertical ($\hat{y}$) directions in the laboratory frame, respectively (referred to as the QPD segmentation axes). GLMT calculation results are compared with the experimental results in terms of the normalized stiffness, $\tilde{k}_{i}=k_{i}/\bar{k}$, where the average stiffness is defined as $\bar{k}=(\bar{k}_{x}+\bar{k}_{y})/2$. For the 300-nm PS spheres presented in Fig. 3, the symbols ($\put(2.5,3){\color{gray}\circle*{7}}\,\,\,$, ${\color{gray}\blacksquare}$) and the gray (solid, dashed) curves denote the experimental and theoretical values of the normalized transverse stiffnesses, $\tilde{k}_x$ and $\tilde{k}_y$, respectively. The orthogonal pair of $\tilde{k}_x$ and $\tilde{k}_y$ shows periodic variation as the input beam polarization angle $\theta_R$ is rotated, which could be inferred from the rotation of the elongated intensity distribution in Fig. 1(b). The average value of the experimentally measured trap stiffness was $\bar{k}_\text{EXP}$ = 59 pN/$\mu$m, and the GLMT calculation with the minimum spot size ({\it w$'_{0}$} = 300 nm) predicts $\bar{k}_\text{GLMT}$ = 60 pN/$\mu$m, showing good agreement with the experimental results. From the angular measurements, the stiffness asymmetry is estimated to be $s_{T,\text{EXP}}$ = 0.29, and the GLMT calculation gives $s_{T,\text{GLMT}}$ = 0.28. 

\begin{figure}[t]
\centering
\includegraphics[width=0.6\columnwidth]{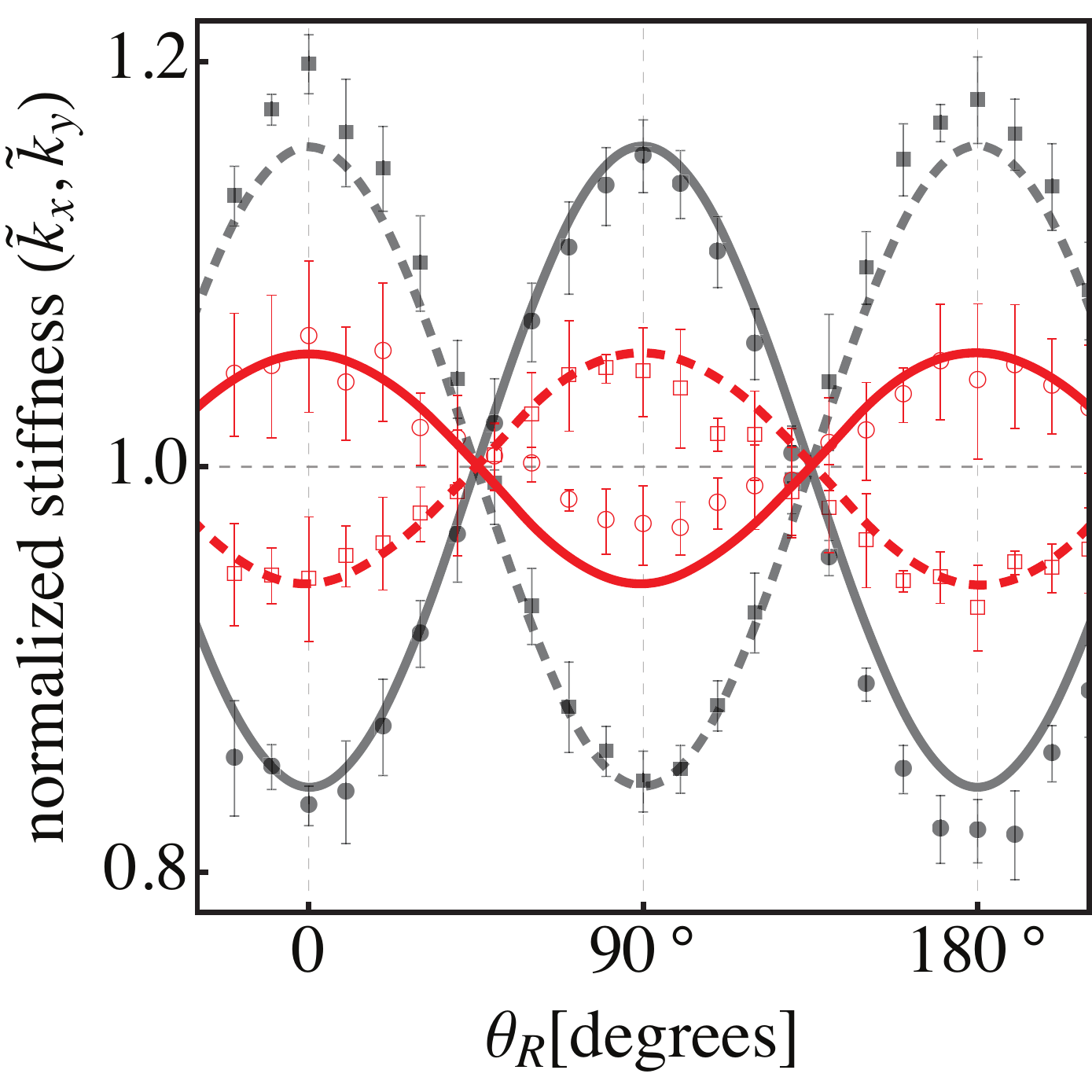}
\caption{Angular dependence of the normalized trap stiffness, $\tilde{k}_x$ and $\tilde{k}_y$. Results of the 300-nm PS sphere: The filled gray (circle, square) symbols and gray (solid, dashed) curves represent the experimental and theoretical values of $\tilde{k}_x$ and $\tilde{k}_y$, respectively. Results of the 1-$\mu$m PS sphere: The empty red (circle, square) symbols and red (solid, dashed) curves represent the experimental and theoretical values of $\tilde{k}_x$ and $\tilde{k}_y$, respectively.}\label{fig3}
\end{figure}

Figure 3 also contains the results of the 1-$\mu$m PS spheres: the symbols ($\put(2.5,3){\color{red}\circle*{7}}\put(2.5,3){\color{white}\circle*{6}}\,\,\,$, ${\color{red}\square}$) and the red (solid, dashed) curves denote the experimental and theoretical values of the normalized transverse stiffnesses, $\tilde{k}_x$ and $\tilde{k}_y$, respectively. We note the inversion of the angular stiffnesses, $\tilde{k}_x$ and $\tilde{k}_y$, between the 300-nm and 1-$\mu$m PS spheres; for the 300-nm PS spheres, $\tilde{k}_x < \tilde{k}_y$ at $\theta_R$ = 0$^{\circ}$ owing to the elongated intensity distribution of the linearly polarized beam, whereas   $\tilde{k}_x > \tilde{k}_y$   in case of the 1-$\mu$m PS spheres with the same polarization configuration. This implies significant EM field redistribution by the presence of a large dielectric particle, even in spherical shape. The average value of the experimentally measured trap stiffness and the theoretical estimation of the 1-$\mu$m PS spheres were ($\bar{k}_\text{EXP}$, $\bar{k}_\text{GLMT}$) = (97, 125) pN/$\mu$m, and the stiffness asymmetry factors were ($s_{T,\text{EXP}}$, $s_{T,\text{GLMT}}$) = (-0.1, -0.12), respectively. Although the experimental results and theoretical estimations show qualitative agreement (sign change of asymmetry factor), there is about 20$\%$ discrepancy in the 1-$\mu$m PS spheres results, which probably originates from the fifth-order Gaussian beam method used in GLMT calculation \cite{Noh2015,Barton19891,Barton19892}. High-order Gaussian beam method provides the focused EM field distribution in the transverse plane with errors of within a few percentages; however, it renders a steeper intensity variation along the axial direction compared to the vectorial diffraction theory \cite{Rohrbach2001}. 

Figure 4(a) shows a demonstration of trap asymmetry control, which proceeds by adjusting the polarization state of the input beam. The polarization state of the input beam is prepared as $\vec{E}^{in}=E_{0}(\text{cos}\theta_{\text{QWP}}\hat{x}+i\text{ sin}\theta_{\text{QWP}}\hat{y})$, where $\theta_\text{QWP}$ is the angle between the  slow axis of the QWP and the horizontal ($\hat{x}$) axis. As the polarization state becomes circular polarization at  $\theta_\text{QWP} = 45^{\circ}$, the strong asymmetry of trap stiffness diminishes for both the 300-nm and 1-$\mu$m PS spheres. In particular, the inverted stiffness asymmetry ($s_\text{T} < 0$) of the 1-$\mu$m PS spheres could also be readjusted to be isotropic with a circularly polarized trap beam, which can be inferred from the rotational symmetry of the isotropic field distribution of a circularly polarized beam and the PS spheres. To explore the trap potential landscape in the transverse plane, we investigated the angular variation in the balanced trap stiffnesses as follows. For each angles of HWP2, the direction of the input beam polarization is rotated to $\theta_{\text{R}}  = 2\theta_\text{HWP2}$, and the QWP is rotated to produce a circularly polarized beam for the given polarization angle: $\theta_\text{QWP}=\theta_\text{R}+45^{\circ}$. Figure 4(b) shows moderate variation in the normalized stiffnesses compared to the previously discussed angular variation in the trap stiffnesses for linearly polarized light (presented as light curves for comparison). The average stiffness for the 300-nm (1-$\mu$m) PS spheres was 59 (98) pN/$\mu$m with the 1$\sigma$-level standard deviation of 0.9 (0.9) pN/$\mu$m, corresponding to fluctuations of 1.4 (1) $\%$, respectively.

\begin{figure}[t]
\centering
\includegraphics[width=0.6\columnwidth]{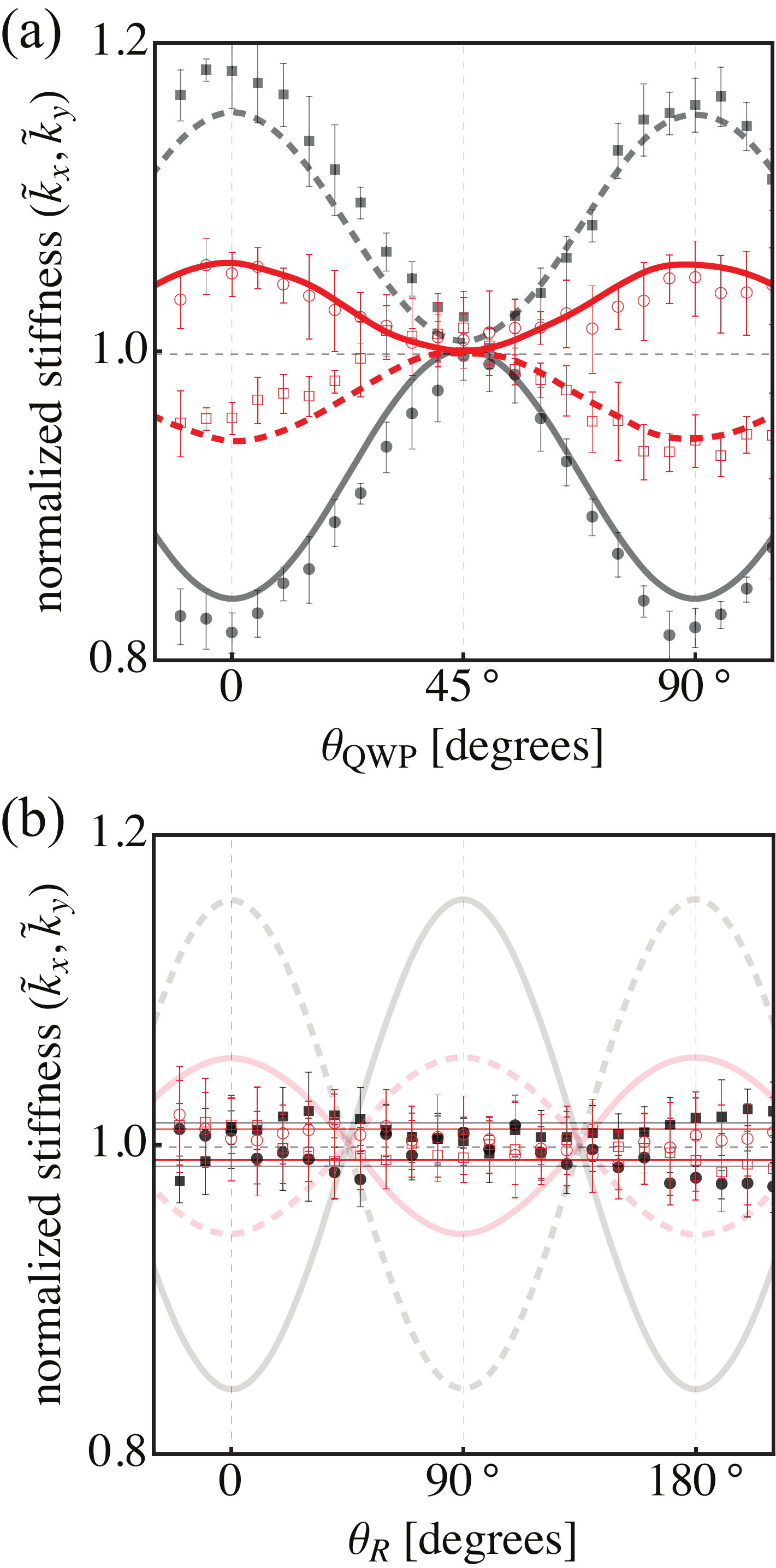}
\caption{Trap stiffness control via the polarization state of trap beam. Results of the 300-nm PS sphere: The  filled gray (circle, square) symbols and gray (solid, dashed) curves represent the experimental and theoretical values of  $\tilde{k}_x$ and $\tilde{k}_y$ respectively. Results of the 1-$\mu$m PS sphere: The empty red (circle, square) symbols and red (solid, dashed) curves represent the experimental and theoretical values of $\tilde{k}_x$ and $\tilde{k}_y$, respectively. (b) Trap stiffnesses, $\tilde{k}_x$ and $\tilde{k}_y$, variation of the balanced potential. Same symbols are used. The horizontal gray (red) lines represent the 1 $\sigma$-level standard deviation of the 300-nm (1-$\mu$m) PS spheres, respectively.}\label{fig4}
\end{figure}

\begin{figure}[!h]
\centering
\includegraphics[width=\columnwidth]{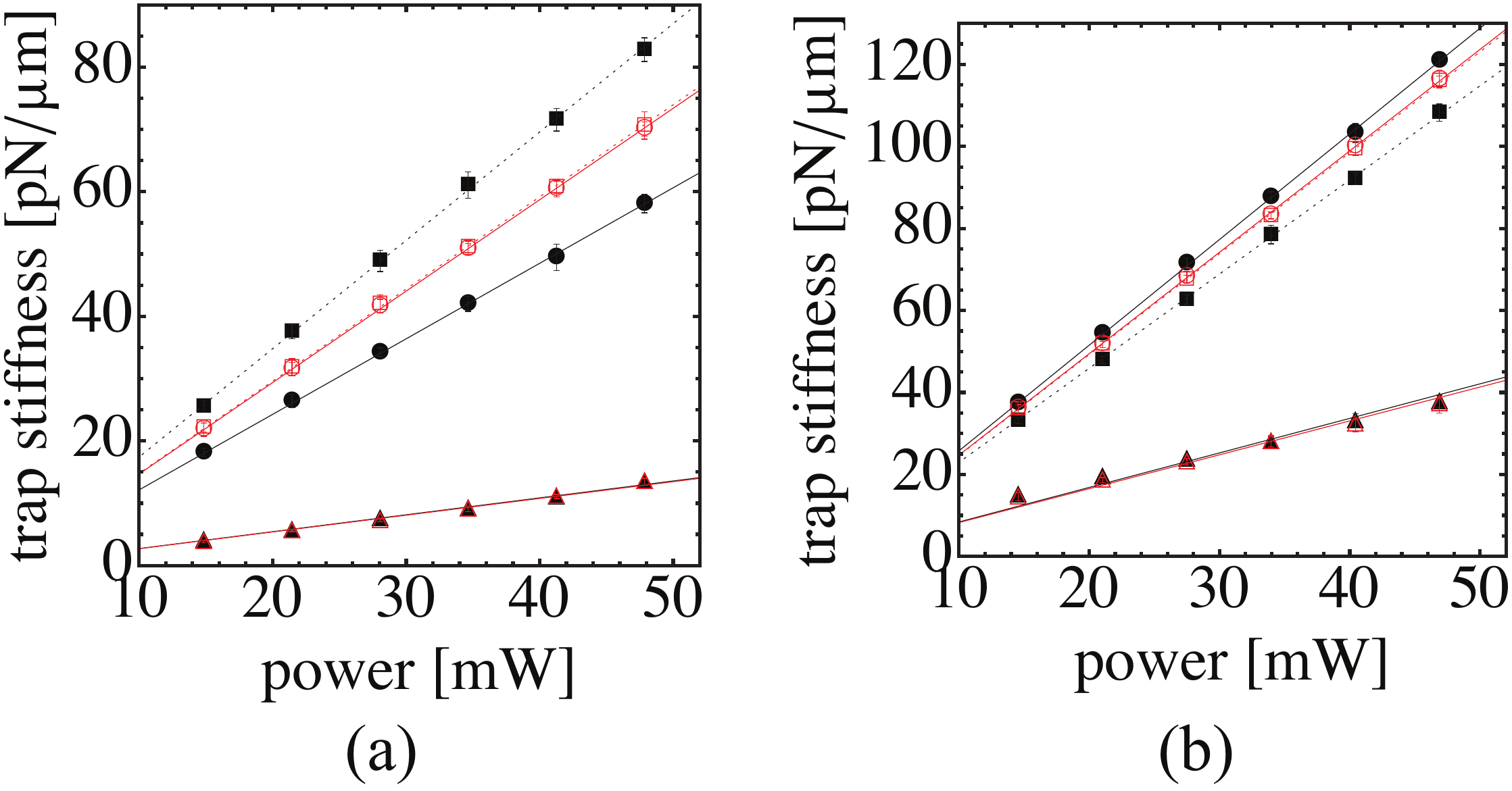}
\caption{Trap stiffness as a function of trap beam power: (a) 300-nm and (b) 1-$\mu$m PS spheres. The (square, circle, triangle) symbols denote $(k_{x}, k_{y}, k_{z})$, respectively, and the (filled, empty) status of the symbols distinguishes the polarization state (LP, CP), respectively. For most data points, the error bars ($\pm 1\sigma$ level) are smaller than the size of the symbols.}\label{fig5}
\end{figure}

Figure 5 shows the linear dependence of trap stiffness on the beam power for linear and circular polarization at a fixed trap depth of 50 $\mu$m from the inner wall of the cover glass. Other measurements (not presented) of trap stiffness as a function of the trap depth over the range of 10--150 $\mu$m show a few percentages of random fluctuations, as is typical for a water-immersion objective lens \cite{Madadi2012,Mahmoudi2011}. The trap stiffness per beam power for each degree of freedom is found by the linear fit of the experimental data: $(k'_{x}, k'_{y}, k'_{z})^\text{LP}$ = (1.21, 1.74, 0.27) pN/$\mu$m/mW and  $(k'_{x}, k'_{y}, k'_{z})^\text{CP}$  = (1.47, 1.48, 0.27) pN/$\mu$m/mW for the 300-nm PS spheres; and   $(k'_{x}, k'_{y}, k'_{z})^\text{LP}$ = (2.58, 2.30, 0.84) pN/$\mu$m/mW and   $(k'_{x}, k'_{y}, k'_{z})^\text{LP}$ = (2.47, 2.46, 0.83) pN/$\mu$m/mW for the 1-$\mu$m PS spheres, where the superscript denotes the polarization states.

\section{Concluding remarks}
The anisotropic field distribution of a highly focused EM fields manifests as dramatically modified optical potential landscape depending on the participating dielectric spheres, i.e., in terms of the sign and magnitude of stiffness asymmetry factor: $s^\text{LP}_{T,\text{EXP}}$ = (0.29, -0.1) for 300-nm and 1-$\mu$m PS spheres, respectively. In case of the adjusted optical trap realized by a simple polarization control, the angular measurements of the orthogonal stiffness pair show less than 1.5$\%$ variation in a statistical sense ($|s^\text{CP}_{T,\text{EXP}}| < 0.03$). Our research translates the polarization state of the trap laser beam, which is responsible for trap potential asymmetry, into a control parameter for tailoring the optical potential landscape of OTs in the transverse directions.

\begin{acknowledgments}
Authors acknowledge the careful reading of the manuscript by Prof. D. Cho and the valuable comments from the optics community. This work was supported by the Basic Science Research Program (No. 2011-0014908) of the National Research Foundation of Korea (NRF), which has been funded by the Ministry of Education, Science, and Technology.
\end{acknowledgments}

%\begin{references}
%\bibliography{BibliographyA3WT}
%\end{references}

%\section*{References}
%\bibliography{BibliographyA3WT}

\end{document}